\documentclass[11pt,a4paper]{article}
\pdfoutput=1
\usepackage{jheppub}

\usepackage{xspace}
\usepackage{epsfig}
\usepackage{amssymb,amsmath}
\usepackage{url,booktabs}
\usepackage{color}

\def\be{\begin{equation}}
\def\ee{\end{equation}}
\def\bea{\begin{eqnarray}}
\def\eea{\end{eqnarray}}
\def\beal{\begin{align}}
\def\eeal{\end{align}}

\newcommand{\met}{E_T^{\rm miss}}
\newcommand{\ptj}{p_T (j)}
\newcommand{\ptjone}{p_T (j_1)}
\newcommand{\ptjtwo}{p_T (j_2)}
\newcommand{\ptl}{p_T (\ell)}
\newcommand{\ptmu}{p_T (\mu)}
\newcommand{\ptel}{p_T (e)}
\newcommand{\ptcut}{p_T^{cut}}

\newcommand{\mneut}{m_{\chi_1^0}}
\newcommand{\mchar}{m_{\chi_1^\pm}}

\relax

\title{Compressed electroweakino spectra at the LHC}

\author[a,b,c]{Pedro Schwaller}
\author[d]{and Jos\'e Zurita}

\affiliation[a]{CERN, Theory Division, \\ CH1211 Geneva 23, Switzerland}
\affiliation[b]{HEP Division, Argonne National Laboratory, \\ 9700 Cass Ave., Argonne IL 60439, U. S. A}
\affiliation[c]{Physics Department, University of Illinois at Chicago, \\ Chicago, IL 60637, U. S. A.}
\affiliation[d]{Institut f{\"u}r Physik (THEP) Johannes Gutenberg-Universit\"at, \\ D-55099, Mainz, Germany}

\emailAdd{pedro.schwaller@gmail.com}
\emailAdd{jose.zurita@uni-mainz.de}

\abstract{In this work, we examine the sensitivity of monojet searches at the LHC to directly produced charginos and neutralinos (electroweakinos) in the limit of small mass splitting, where the traditional multilepton plus missing energy searches loose their sensitivity.
We first recast the existing 8~TeV monojet search at CMS in terms of a SUSY simplified model with only light gauginos (winos and binos) or only light Higgsinos. The current searches are not sensitive to MSSM-like production cross sections, but would be sensitive to models with 2~-~20~times enhanced production cross section, for particle masses between 100~GeV and 250~GeV. 
Then we explore the sensitivity in the 14~TeV run of the LHC. Here we emphasise that in addition to the pure monojet search, soft leptons present in the samples can be used to increase the sensitivity. Exclusion of electroweakino masses up to 200~GeV is possible with 300~fb$^{-1}$ at the LHC, if the systematic error can be reduced to the 1\% level. Discovery is possible with 3000~fb$^{-1}$ in some regions of parameter space. }


\arxivnumber{1312.7350}

\begin{document}
\begin{flushright}
CERN-PH-TH/2013-319  \\ MITP/13-084
\end{flushright}
\maketitle

\section{Introduction}
\label{sec:intro}

The discovery of a particle with a mass of about 125~GeV~\cite{LHCHiggs}, consistent with the Standard Model (SM) Higgs boson, dawns a new era in particle physics. After such an achievement, now the attention is turned to understand if this resonance is indeed the SM Higgs boson, or if there is room for Beyond the Standard Model (BSM) phenomena.

Several open issues require new physics to be introduced at some scale. Among those, the hierarchy problem and dark matter suggest that new physics as light as the weak scale could be present. A weakly interacting, neutral and stable particle with mass of order of the weak scale is among the leading candidates for dark matter, while most solutions to the hierarchy problem require weakly interacting partners for the electroweak gauge bosons. 

This strongly motivates a search for weakly-interacting new physics, which, in contrast to strongly coupled new physics, can still be present at or very close to the weak scale. 

In the minimal supersymmetric standard model (MSSM), the superpartners of the weak gauge and Higgs bosons play both an important role in the solution of the hierarchy problem and in providing a candidate for dark matter. We will therefore adopt the MSSM as our ``BSM benchmark'' model.  A typical MSSM signature is given by jets plus a large amount of transverse missing energy ($\met$), originating from gluino and/or squark production, and the subsequent chain decays. Generically these decay chains end with the lightest supersymmetric particle (usually the neutralino) which constitutes a very good dark matter candidate. Current constraints MSSM superpartners can roughly be summarised as follows: Gluinos and squarks of the first and second generation have to be heavier than about a TeV~\cite{ATLAS:gluinos,CMS:gluinos}, stops heavier than 500-700 GeV~\cite{ATLAS:stops,CMS:stops} and electroweakinos (charginos and neutralinos) heavier than 200-300 GeV~\cite{ATLAS:ewkinos,CMS:aro}.

However, these constraints do often rely on certain simplified assumptions that can be relaxed. For instance, if the mass splittings between sparticles is small, then the amount of missing energy as well as the transverse momentum of the associated jets is reduced, and many of these searches become severely less sensitive to the new physics.

In this article we investigate the possibility to use  monojet plus $\met$ searches to look for chargino pair production decaying into neutralinos, leptons and neutrinos (and possibly more jets), in the case where there is a small mass gap between the chargino and the neutralino. 
In the particular context of the MSSM such a spectrum can be obtained by taking, for instance, the gaugino soft masses $M_1,~M_2$ of the order of 100 GeV, while $\mu\sim 1$ TeV. The spectrum then contains two light neutralinos $\chi^0_{1,2}$ and one chargino $\chi^\pm_{1}$ which are gaugino-like and whose mass splitting is controlled by $M_2-M_1$, and heavier states at the scale $\mu$ which are Higgsino-like. 
In the context of Natural SUSY~\cite{naturalsusy} the opposite limit is achieved, namely $\mu \sim {\cal O} (100)~\rm{GeV}$ and $M_1, M_2 \gtrsim 1$ TeV. Qualitatively the same spectrum is obtained, but with the light (heavy) states being Higgsino (gaugino)-like. Such an scenario is also motivated from the fact that in the MSSM the minimization of the Higgs potential requires $\mu$ of the order of the electroweak scale, to avoid a large fine-tuning.

When the mass splittings are small, visible decay products from the decays of the heavier $\chi_2^0$ and $\chi_1^\pm$ states become too soft and do not pass the trigger requirements employed in most BSM searches. Therefore we will require an additional hard jet from ISR radiation to boost the missing energy, such that at least in principle, the signal can be recorded. This is the basic idea behind the monojet search for dark matter. 

We will analyse both the sensitivity of the existing 8~TeV monojet searches and present projections for the sensitivity of the future 14~TeV high and very high luminosity runs of the LHC. Furthermore we will attempt to improve the sensitivity of the monojet searches to MSSM-like scenarios by using soft leptons that are likely to be present in the samples. 

This paper is organized as follows.  In Sec.~\ref{sec:status}, we review the phenomenology of the electroweakino sector in the MSSM, discuss the current exclusions from collider experiments and bounds from dark matter. In Sec.~\ref{sec:ISR} we elaborate on our strategy and discuss the parameter space under study.
In Sec.~\ref{sec:CMS} we explain our setup and validate our Monte Carlo simulation against the results of the CMS analysis. In Sec.~\ref{sec:8TeV} we recast the current experimental results for the 8 TeV LHC, while in Sec.~\ref{sec:14TeV} we present the reach of the 14 TeV LHC, for two benchmark luminosities of 300 and 3000 fb$^{-1}$. We conclude in Sec.~\ref{sec:conclu}.

\section{Phenomenology and Limits on Electroweakinos}
\label{sec:status}
%
Charginos and neutralinos are the superpartners of the weakly interacting bosonic fields in the SM. The partners of the electroweak gauge bosons, the \textit{winos} and \textit{binos}, and the partners of the two MSSM Higgs doublets, the \textit{Higgsinos}, mix under the influence of electroweak symmetry breaking. The chargino mass matrix is given by~\cite{Martin:1997ns}
\begin{align}
	{\cal M}^\pm_\chi &= 
	\begin{pmatrix}
		M_2 & \sqrt{2} \sin\beta M_W \\
		\sqrt{2} \cos\beta M_W & \mu
	\end{pmatrix},
\end{align}
where $M_2$ and $\mu$ are the supersymmetry breaking wino and Higgsino masses, respectively, $M_W$ is the $W$-boson mass and $\tan\beta=v_u/v_d$ is the ratio of the vacuum expectation values of the two Higgs doublets. 
The neutral states mix according to the neutralino mass matrix, given by
\begin{align}
	{\cal M}^0_\chi & =
	\begin{pmatrix}
		M_1 & 0 & - c_\beta s_W M_Z & s_\beta s_W M_Z \\
		0 & M_2 & c_\beta c_W M_Z & -s_\beta c_W M_Z \\
		- c_\beta s_W M_Z & c_\beta c_W M_Z & 0 & -\mu \\
		s_\beta s_W M_Z & -s_\beta c_W M_Z & -\mu & 0
	\end{pmatrix},
\end{align}
where $s_W$ and $c_W$ are the sine and cosine of the Weinberg angle, $M_Z$ is the mass of the $Z$-boson, and the bino mass $M_1$ appears in addition to the wino and Higgsino masses. 

After the diagonalization, the mass eigenstates are labelled as $\chi_{1,2}^\pm$ and $\chi^0_{1,2,3,4}$ with masses increasing with the index. The lightest neutralino, $\chi^0_1$, is usually the lightest supersymmetric particle, and therefore stable and a candidate for dark matter, provided that $R$-parity is conserved. 

A quick inspection of the mass matrix reveals that the mixing of gauginos and Higgsinos is controlled by a weak scale parameter. Therefore, if either the gaugino or the Higgsino mass is much larger than the weak scale, the mixing is suppressed and one obtains a simpler model with two neutral and one charged state, and a small number of parameters. 

Simplified models~\cite{Alves:2011wf} have successfully been invoked by the LHC experiments to present the results of searches for new physics in a more model independent way, 
and to facilitate the interpretation of the results in different BSM scenarios. In this spirit, we will perform the majority of our analyses in the limiting cases where either $\mu \gg M_1, M_2\sim M_Z$, such that only the electroweak gauginos are present at the weak scale, or where $M_1, M_2 \gg \mu \sim M_Z$, such that only Higgsinos are light. For definiteness, the heavy mass parameters will be set to 1~TeV. We will furthermore assume that other MSSM degrees of freedom are heavy enough to not contribute to chargino and neutralino production. Since third generation colored superpartners, sleptons and MSSM Higgs bosons might still be relatively light, this makes our limits conservative. 

The LEP experiments have searched for charginos and neutralinos produced in the process $e^+ e^- \to \chi \chi'$. Since $\chi_1^0$ does not decay and does not interact with the detector, no direct limit can be obtained from this process. However the precise measurement of the $Z$-boson width imposes a limit of $m_{\chi_1^0} \gtrsim 45$~GeV, unless the neutralino has a very small (or zero coupling) to the $Z$, see e.g~\cite{Dreiner:2009ic}. Searches for heavier charginos and neutralinos were essentially limited by the maximal LEP II center of mass energy of $209$~GeV. The resulting limits lie between $91.9$~GeV and $103.5$~GeV~\cite{Beringer:1900zz}, depending on the details of the production and decay processes. 

In the simplest scenarios, the lightest chargino decays via $\chi^\pm_1 \to W^{\pm(*)} \chi^0_1$, emitting a $W$ boson, while the second lightest neutralino decays through $\chi^0_2 \to Z^{(*)} \chi^0_1$. Depending on the mass difference, the $W$ and $Z$ bosons in these processes can be off-shell. In principle longer and more complicated decay chains can occur, in particular if other light states, for example sleptons, are involved. For the region of parameter space we are interested in, these two processes are however sufficient to understand the phenomenology.\footnote{In addition to $W-$ and $Z$-bosons, the Higgs boson can also appear in these decays. For small mass splittings the decays involving an off-shell Higgs boson are however suppressed by the small width of the Higgs. } 

\

At hadron colliders, the actual center of mass energy of a collision is not known, and furthermore backgrounds from QCD processes make it difficult to identify rare processes such as the production of a pair of charginos or neutralinos. Therefore current searches at hadron colliders have focussed on final states where the intermediate $W$ and $Z$-bosons decay leptonically. The resulting three and four lepton final states, e.g. from processes of the form
\begin{align}
	p p &\to \chi^0_2 \chi^\pm_1 \to W^{\pm} Z \chi_1^0 \chi_1^0 \to \ell^\pm \ell^+ \ell^- \met \,,\\
	p p &\to \chi^0_2 \chi^0_2 \to Z Z \chi_1^0 \chi_1^0 \to \ell^+\ell^- \ell^+ \ell^- \met\,,
\end{align}
have reasonably low backgrounds, such that successful searches are possible even in the busy environment of a hadron collider. 

\begin{figure}
\begin{center}
	\includegraphics[width=.6\textwidth]{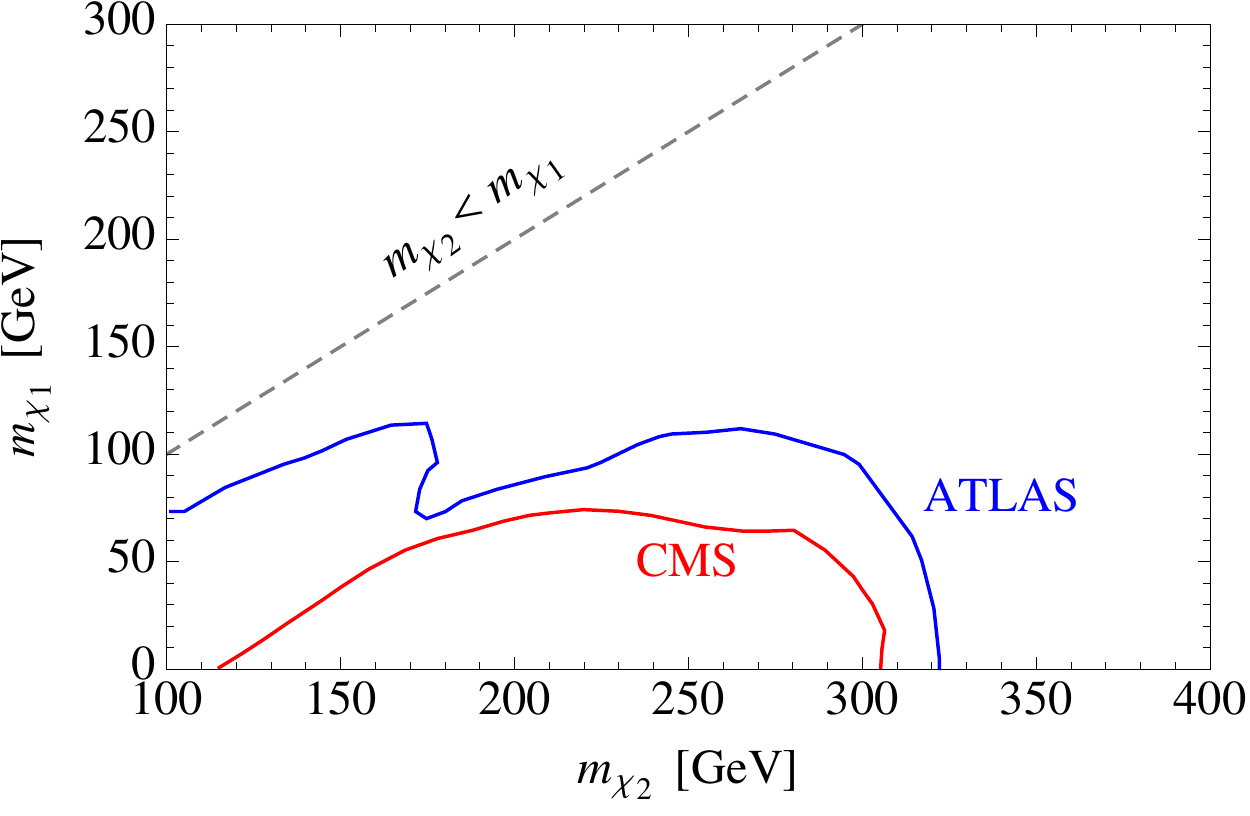}
	\caption{ATLAS (blue) and CMS (red) limits on electroweakly produced charginos and neutralinos decaying via $W$- and $Z$-bosons, assuming $m_{\chi_1^\pm} = m_{\chi_2^0}$.  \label{Fig:current limits}}
\end{center}
\end{figure}

The most recent limits on directly produced charginos and neutralinos from the LHC experiments were obtained using $20.7$~fb$^{-1}$ of $8$~TeV data at ATLAS~\cite{ATLAS:2013rla} and $19.5$~fb$^{-1}$ at CMS~\cite{CMS:2013dea}. The strongest constraints are obtained in the $3$ lepton $+\met$ channel, and the interpretation of the limits in the neutralino-chargino mass plane, assuming $m_{\chi_2^0}=m_{\chi_1^\pm}$, is reproduced in Fig.~\ref{Fig:current limits}. 

While the LHC experiments were able to constrain electroweakino masses up to 320~GeV in the limit of massless LSP, the existing search techniques fail when the mass differences between the lightest and the heavier charginos and neutralinos become small, since then the leptons from the decays become too soft to be detected. In this regime, the relevant decays are essentially three body decays $\chi' \to \ell \ell' \chi^0_1$, such that the lepton energy, in the rest frame of the decaying particle, is roughly bounded by $(m_{\chi'} - m_{\chi_1^0})/2$. 

Standard searches for multi lepton plus $\met$ signals rely on charged lepton triggers that require $p_{T,\ell}\gtrsim 20$~GeV for the hardest lepton\footnote{The CMS trilepton search requires $p_{T,\ell}\geq 20$~GeV for the hardest lepton, and $p_{T,\ell} \geq 10$~GeV for all other leptons. ATLAS is using several different triggers requiring a single lepton with $p_T\geq 25$~GeV, a pair of leptons with $p_T\geq 14$~GeV or one lepton with $p_T\geq 18$~GeV and a second lepton with $p_T\geq 10$~GeV.  }. 
Therefore most of these searches start loosing steam when the mass difference drops below $\sim 50$~GeV, resulting in a loss of sensitivity in the regime where $m_{\chi_2} - m_{\chi_1} \lesssim 50$~GeV. It is exactly in this region where light charginos and neutralinos, or other weakly coupled new physics, might be hiding\footnote{For other strategies to search for charginos and neutralinos, see Ref.~\cite{Cabrera:2012gf,Han:2013kza,Buckley:2013kua}.}. 

In the next section, we will outline our strategy to search for new physics with almost degenerate spectra, and compare it to other proposals. Before going there however, we have to address the question of what happens to the lightest neutralino. While we do not necessarily insist that $\chi_1^0$ constitutes all of the dark matter in the universe\footnote{For a recent discussion see e.g.~\cite{Cheung:2012qy}.}, it is rather important to make sure that its relic density is not too large, since otherwise it would lead to disagreement with the observed universe. 

Higgsinos annihilate very efficiently since they carry electroweak charges, and the same is true for winos. Therefore scenarios with $\mu\ll M_1, M_2$ or cases with $M_2 \ll M_1,\mu$ do not pose any problems for cosmology. However if the LSP is mostly bino, i.e. $M_1 \ll M_2, \mu$, annihilation becomes very inefficient, and the relic density can be problematic. While one could impose the relic density as a constraint on the parameter space, in the spirit of simplified models we will not do this here in order not to cut away regions of parameter space that might be interesting from a collider perspective. Instead we note that there are several alternate possibilities to avoid a too large relic density in this case. 

First, if $\chi_1^0$ is not absolutely stable, but just long lived enough to escape the detectors, the relic density constraint is satisfied while the collider phenomenology is unchanged. Such a scenario can happen for example if R-parity is broken very weakly (see e.g~\cite{Barbier:2004ez}) , or if the neutralino is not the actual LSP and decays e.g. into gravitinos  (see e.g~\cite{Hall:2013uga}). Another possibility is a non-standard cosmological history, for example late decaying particles can inject additional entropy after $\chi_1^0$ freezes out, such that its relic density is diluted (see e.g ~\cite{McDonald:1989jd}). 

\

\section{Search strategy and parameter space}
\label{sec:ISR}
%
%
When a pair of charginos or neutralinos is produced at rest, both the missing transverse energy $\met$ and the lepton momentum $\ptl$ are bounded to be of the same order as the mass gap between the heavier and lighter states. To see this, consider a simple two body decay involving one massless particle, in the rest frame of the decaying particle. The massless particles momentum is then given by
\begin{align}
	p & = \frac{m_2^2 - m_1^2}{2 m_2} \approx \Delta m
\end{align}
with $m_2 = m_1 + \Delta m$ and assuming $\Delta m \ll m_1$. Since charginos and neutralinos typically experience three body decays, the momentum is further reduced. If there is no additional radiation in the event, the total $\met$ in the event, i.e. the vector sum of the transverse lepton momenta, is approximately bounded by $2\Delta m$. Once the $\met$ and the lepton momenta fall below the trigger requirements, the searches loose sensitivity. 

Additional jets from ISR can be used to enhance the detectability of this signal. First, the invisible particles can now recoil against the jet, such that $\met \sim \ptj$, and the signal can at least be triggered. In addition, also the lepton momenta receive a part of the boost, which makes them more likely to pass the leptonic triggers. In~\cite{Gori:2013ala}, the extra ISR jet has been used to increase the sensitivity of the trilepton search. Note however that it is not possible to go to arbitrary small mass gaps, since the boost only acts multiplicatively on the small lepton momenta. 

Here instead our aim is to fully close the gap down to the degenerate limit, using only the monojet plus missing energy signal as trigger. The possibility to search and constrain dark matter models using this signature has been suggested in~\cite{Goodman:2010yf,Bai:2010hh,Goodman:2010ku} and has been successfully employed to constrain effective dark matter models by the ATLAS and CMS collaborations~\cite{Chatrchyan:2012me,CMS:rwa,ATLAS:2012ky,ATLAS:2012zim}. The importance of ISR radiation for SUSY searches has also been emphasised and exploited, for electroweakino production at the Tevatron~\cite{Gunion:1999jr}, and in~\cite{Alwall:2008ve,Carena:2008mj,Alwall:2008va,Izaguirre:2010nj,LeCompte:2011cn,LeCompte:2011fh,He:2011tp,Drees:2012dd,Alvarez:2012wf,Dreiner:2012gx,Bhattacherjee:2012mz,Dreiner:2012sh,Arbey:2013iza} in the context of strong SUSY production. A first application to direct electroweak production of Higgsinos has also appeared recently~\cite{Han:2013usa}. 

For the pair production of electroweakinos in association with an additional jet it is worth noting that the additional factor of $\alpha_s$ is partially compensated by the larger gluon PDFs at the LHC. The reason is that the extra jet will allow contributions from quark gluon initial states, while the pure pair production process originates from quark anti-quark initial states at the tree level, and the anti-quark PDF is suppressed at a proton proton collider.  

Beyond the pure monojet signature, some soft leptons remain in the samples. The CMS monojet study applies a lepton veto, that is, reject events  with $\ptl > 10$~GeV. However, events with leptons can still pass this cut: CMS identifies isolated muons down to $\ptl \geq 3$~GeV, while electrons can be reconstructed down to $5-10$~GeV.

It was first suggested in~\cite{Giudice:2010wb} (see also~\cite{Rolbiecki:2012gn}) to utilise these additional leptons in SUSY searches. In our 14~TeV analysis we will further divide the monojet signal region into regions with zero, one or two soft leptons, and show that this can improve the sensitivity compared to the pure monojet search. 
Apart from the soft lepton analysis, we will stick as close as possible to the existing CMS search. 

To evaluate the sensitivity of the search to degenerate electroweakinos, we will perform Monte Carlo simulations for a set of representative points in the MSSM parameter space, using the spectrum calculator SUSPECT2~\cite{Djouadi:2002ze}. All scalar superpartners and the gluinos are set to be at the multi-TeV scale. 

For the light gaugino case, we set $\mu=1$~TeV and vary $M_1, M_2$ in the 100-250 GeV range, with $ |M_2 - M_1| \le $ 30 GeV. In this way we obtain scenarios with mass splitting ranging from 0-40~GeV. Here $m_{\chi^\pm_1} \approx m_{\chi^0_2}$, therefore we will present our results as functions of $m_{\chi^0_1}$ and $m_{\chi^\pm_1}$ only. 

For the case of light Higgsinos, $M_1 = M_2=1$~TeV, we use $\mu$ as the only free parameter and vary it between 100~GeV and 250~GeV. The mass splittings are of order $m_W^2/M_{1,2} \lesssim 10$~GeV and almost independent of $\mu$, but could be smaller if $M_{1,2}$ was increased further. Since the effects of varying mass splittings are already covered by the light gaugino case, we will keep $M_{1,2}$ fixed here. 

\begin{figure}[htp]
\begin{center}
\includegraphics[width=.6\textwidth]{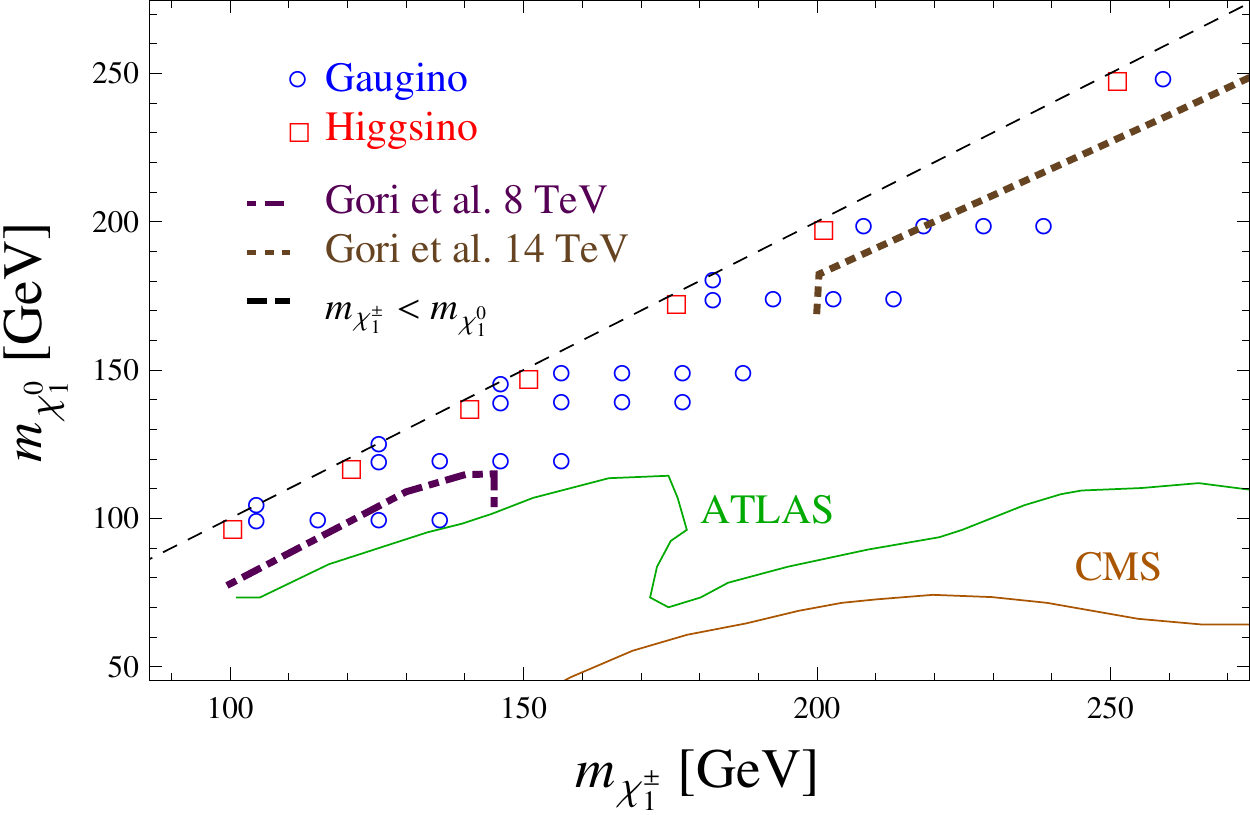}
\caption{Location of the simulated MSSM parameter points in the chargino-neutralino mass plane. Blue circles denote light gaugino scenarios, while the red squares are the Higgsino parameter points. Also shown are the existing constraints from the 8~TeV analyses of ATLAS and CMS, as well as the expected sensitivities from the boosted trilepton search of Ref.~\cite{Gori:2013ala}.
}
 \label{fig:parameterpoints}
\end{center}
\end{figure}
The location of our parameter points in the chargino-neutralino mass plane is shown in Fig.~\ref{fig:parameterpoints}. All points lie in a region that is currently not probed by the LHC experiments, and that will remain difficult to probe in the future, although some of them are in reach of the boosted trilepton search proposed in Ref.~\cite{Gori:2013ala}.

\section{CMS monojet analysis and Monte Carlo validation}
\label{sec:CMS}
CMS has performed a search for new physics in the monojet plus $\met$ channel in 19.5~fb$^{-1}$ of data from the 8~TeV run of the LHC~\cite{CMS:rwa}. In this section we summarise the CMS analysis and validate our Monte Carlo setup by comparing our background model to the data. 

Events are recorded using a set of triggers that require $\met > 120$~GeV and a jet with $\ptjone > 80$~GeV within $|\eta| < 2.6$. Furthermore a set of \textit{preselection cuts} is employed which require $\met > 200$~GeV and $\ptjone > 120$~GeV within $|\eta| < 2.4$.

The main background processes are $W/Z$+jets production, $t \bar{t}$, single top and QCD multijet backgrounds. Of those, the $Z$+jets background with the $Z$ decaying to neutrinos is irreducible. A series of cuts is employed to reduce the other backgrounds:
\begin{itemize}
\item Veto on a third-jet: $N_{jet} (p_T > 30~{\rm GeV}) \le 2$.
\item Dijet angular cut: $\Delta \phi (j_1, j_2) < 2.5$.
\item Lepton veto: Events with isolated muons or electrons with $\ptl > 10$ GeV, or with reconstructed taus with $p_T (\tau) > 20$ GeV and $\lvert \eta (\tau) \rvert < $  2.3 are rejected.
\end{itemize}
The third jet cut together with the angular cut very efficiently suppress the $t\bar{t}$ and multijet backgrounds, while the lepton veto significantly reduces the backgrounds from leptonic $W$ decays. 
The analysis is then divided into seven $\met$ bins with $\met > 250$, 300, 350, 400, 450, 500, 550~GeV, and a limit is derived from the bin that gives the largest significance. 

\

To validate our Monte Carlo setup, we generate the dominant $(W\to \ell \nu)$+jets and $(Z\to\nu\nu)$+jets backgrounds as well as the $t\bar{t}$ and the $(Z\to\ell\ell)$+jets backgrounds. The single top and QCD multijet backgrounds are less relevant are therefore neglected here. 

Background events are generated using MadGraph5~\cite{Alwall:2011uj} and then passed to Pythia~6~\cite{Sjostrand:2006za} for parton showers and hadronization and to Delphes~2~\cite{Ovyn:2009tx} to perform a fast detector simulation. Jets are clustered~\cite{Cacciari:2011ma} using the anti-$k_t$ algorithm~\cite{Cacciari:2008gp} with $R=0.5$. For the $W/Z$+jets backgrounds we generate parton level events with up to two jets that are matched to the parton shower using the MLM-scheme~\cite{Mangano:2002ea} as implemented in MadGraph5. Since $t\bar{t}$ events already contain a large number of jets we have only generated events with up to one additional jet at the parton level. For the detector simulation, we have adjusted the Delphes~2 CMS input card to mimic the CMS analysis as closely as possible. Furthermore we have lowered the $\ptmu$ threshold to 5~GeV to allow for \textit{soft muons} in our sample. 
\begin{figure}[t]
\begin{center}
\begin{minipage}[b]{0.47\linewidth}
\begin{center}
\includegraphics[width=1\textwidth]{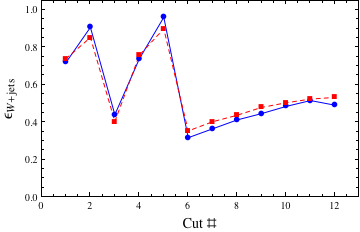}
\newline
(a)
\end{center}
\end{minipage}
\hspace{0.5cm}
\begin{minipage}[b]{0.47\linewidth}
\begin{center}
\includegraphics[width=1\textwidth]{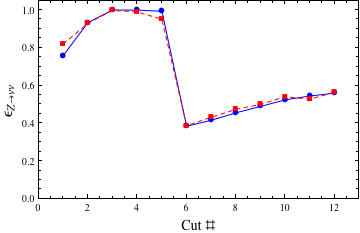}
\newline
(b)
\end{center}
\end{minipage}
\caption{Efficiencies for CMS (solid blue) and our MadGraph simulation (dashed red) for (a) $(W\to\ell\nu_\ell)+{\rm jets}$ and (b) $(Z \to \nu \bar{\nu})$ + jets processes, for the 8 TeV LHC. Cuts as described in the text. }
 \label{fig:eff8TeV}
\end{center}
\end{figure}

We have reproduced several kinematic distributions presented in Ref.~\cite{CMS:rwa} with good accuracy. To further compare our event samples with the CMS data we define the \emph{efficiency} of a cut as
\be
\epsilon_i = \frac{N_{i+1}}{N_i} \, ,
\ee
where $N_i$ is the number of events after the i-th cut, with the numbers referring to the cuts listed in Tab.~1 of~\cite{CMS:rwa} and $i=1$ corresponding to the preselection cut. The efficiencies for $W$ + jets and $(Z \to \nu \nu)+{\rm jets}$ are compared to those in the CMS study in Fig.~\ref{fig:eff8TeV}. We see that we have an accurate description of both  processes, with discrepancies at most at the few percent level. 

Before moving to the next section, it is worth noting that the dominant $(Z\to\nu\nu)+{\rm jets}$ and $(W\to\ell \nu_\ell)+{\rm jets}$ backgrounds can be accurately modelled using data driven methods, as is already done in both the CMS study discussed here as well as in the ATLAS monojet search. This efficiently reduces the theory uncertainty on the backgrounds, and will allow for small systematic errors in high luminosity studies at the 14~TeV LHC. 
\section{Recast of 8 TeV data}
\label{sec:8TeV}

\begin{figure}[!htp]
\begin{center}
\includegraphics[width=.45\textwidth]{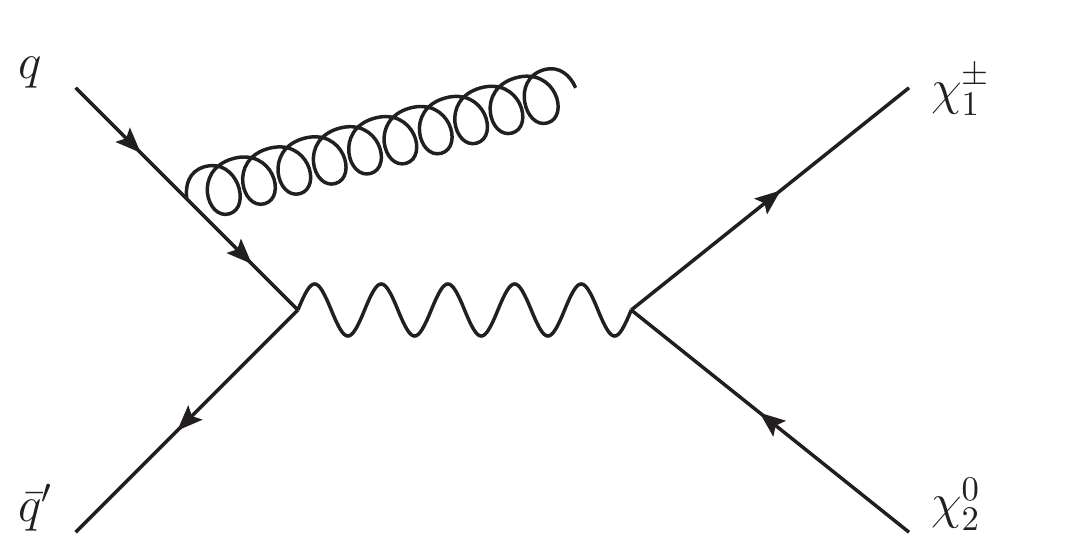}
\hspace*{1cm}
\includegraphics[width=.45\textwidth]{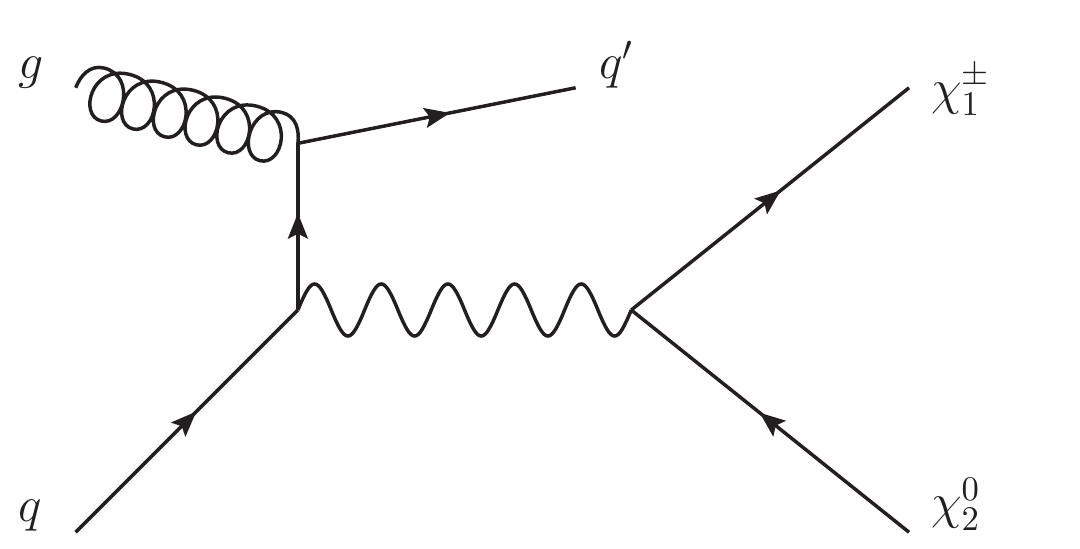}
\caption{Two processes that contribute to electroweakino + jet production at leading order, and that were computed at the parton level using MadGraph/MadEvent.  }
 \label{fig:diagram}
\end{center}
\end{figure}

We generate the processes $p p \to \chi \chi' + 1,2$~jets for the MSSM parameter points discussed in Sec.~\ref{sec:ISR} using the MSSM implementation in MadGraph5. As for the background processes, we use MLM matching up to two additional jets and the fast detector simulation Delphes~2, and $\chi, \chi' = \chi^0_1, \chi^0_2, \chi_1^\pm$. Two of the parton level diagrams that are simulated in MadGraph5 are shown in Fig.~\ref{fig:diagram}, highlighting the different initial states, $q\bar{q}$ and $qg$, $\bar{q}g$, that contribute at leading order (LO). Charginos and neutralinos are then decayed in Pythia. 

We note that the $2 \to 3$ signal processes are only known at LO in perturbation theory, except for the case of a neutralino pair, which was computed recently~\cite{Cullen:2012eh}. One could try to estimate the K-factor using PROSPINO~\cite{Beenakker:1996ed}, which gives the $2 \to 2$ inclusive process. For our range of masses the K-factor ranges between 1.3-1.5, depending on the particular process~\cite{Beenakker:1999xh,Fuks:2012qx}. However, assuming that the $2 \to 2$ and $2 \to 3$ processes have the same K-factor is not a well justified approximation, and it was indeed shown in Ref.~\cite{Cullen:2012eh} that the K-factors are not constant over the phase-space. This can be easily understood from the fact that at NLO the $gg$ channel opens up in addition to the LO production modes, whereas this channel does not contribute to the inclusive $2\to 2$ process at NLO. The suppression of the $gg$ channel by one power of the strong coupling constant is partially compensated by the dominance of the gluon parton distribution function at the LHC. For the partonic channels that are open at LO, Ref.~\cite{Cullen:2012eh}  estimates a K-factor of about 2.3, very different from the one for the inclusive $2\to 2$ process. Due to the discrepancy between both estimations we prefer to be on the conservative side, and hence we will not apply any K-factors to our simulated samples. Given the situation described here, it would be utterly necessary to extend the analysis of Ref.~\cite{Cullen:2012eh} to the other signal processes.

\begin{figure}[!htp]
\begin{center}
\includegraphics[width=.5\textwidth]{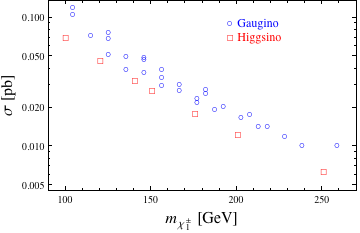}
\caption{Cross section for all electroweakino processes after preselection cuts, as a function of $m_{\chi^{\pm}_1}$. Blue circles (red squares) correspond to the gaugino (Higgsino) case. In the gaugino case multiple points for each $\mchar$ correspond to different values of $m_{\chi^0_1}$, i.e. different mass splittings. }
 \label{fig:char1}
\end{center}
\end{figure}

In Fig.~\ref{fig:char1} we show the total signal cross sections for our parameter points after preselection cuts as a function of $\mchar$.  The cross sections are ${\cal O} (10-100)$ fb and one sees that they  strongly depend on $m_{\chi^{\pm}_1}$. Note that the Higgsino cross section are a factor of two smaller than the gaugino cross section for a similar chargino mass. Given the size of the cross sections, it is obvious that for the 8 TeV LHC dataset statistics will be a limiting factor: before applying any further cuts one is left with  ${\cal O} (100-1000)$ events. 

For a given signal point, we estimate the significance $\sigma$ as the number of signal events $S$ divided by the uncertainty on the background $\Delta B$: 
\be\label{eq:significance}
\sigma = S / \Delta B \, .
\ee
For our recast of the $8$~TeV CMS analysis, $\Delta B$ will be directly taken from the \emph{observed} limit quoted in the CMS paper (which also accounts for fluctuations in the data). However it is useful to decompose $\Delta B$ into statistical and systematic contributions, in order to identify the limiting factors of the analysis. Therefore, we write
\be\label{eq:DeltaB}
\Delta B = \sqrt{\sum_i [B_i + (\beta_i B_i)^2 ]} \, .
\ee
Here, $B_i$ is the number of background events of the $i$-th background process ($\sum_i B_i = B$). For $\beta_i = 0$ this reduces to the well known $\Delta B = \sqrt{B}$ estimate of the statistical error. The factors $\beta_i$ parameterise the systematic errors in the different channels, and we combine systematic and statistical errors in quadrature. In the limit of infinite luminosity we have $\sigma \approx S/(\beta B)$ (for $\beta_i=\beta$) which is a limiting factor for the analysis~\footnote{We note that in the CMS analysis the uncertainty due to the background fluctuations in the control sample is considered as part of the systematic error. Hence this effect is included into the coefficient $\beta_i$. }. 

As pointed out in~\cite{Giudice:2010wb}, there is no shape difference between signal and backgrounds (the only difference being that the signal has a harder $\met$ spectrum), and thus systematic errors can render the identification of the signal very challenging. A direct inspection of the CMS analysis shows that indeed one major component of the error is due to systematics. 
In order to not introduce additional errors, we will therefore closely follow the CMS analysis, and take the $\Delta B$ directly from the observed upper limit that is reported in Tab.~8 of Ref.~\cite{CMS:rwa}. Using our formula for $\Delta B$, we can also estimate the factors $\beta_i$ for the different analysis bins, and find that the total systematic error ranges from about $\beta =4.7\%$ for the lowest $\met$ bin to $\beta=15\%$ for the highest $\met$ bin. Here we have assumed that $\beta_i= \beta$. While this is not a valid assumption in general, the two leading backgrounds have similar systematics, and the remaining backgrounds can almost be neglected. 

\begin{figure}[!htp]
\begin{center}
\begin{minipage}[b]{0.47\linewidth}
\begin{center}
\includegraphics[width=1\textwidth]{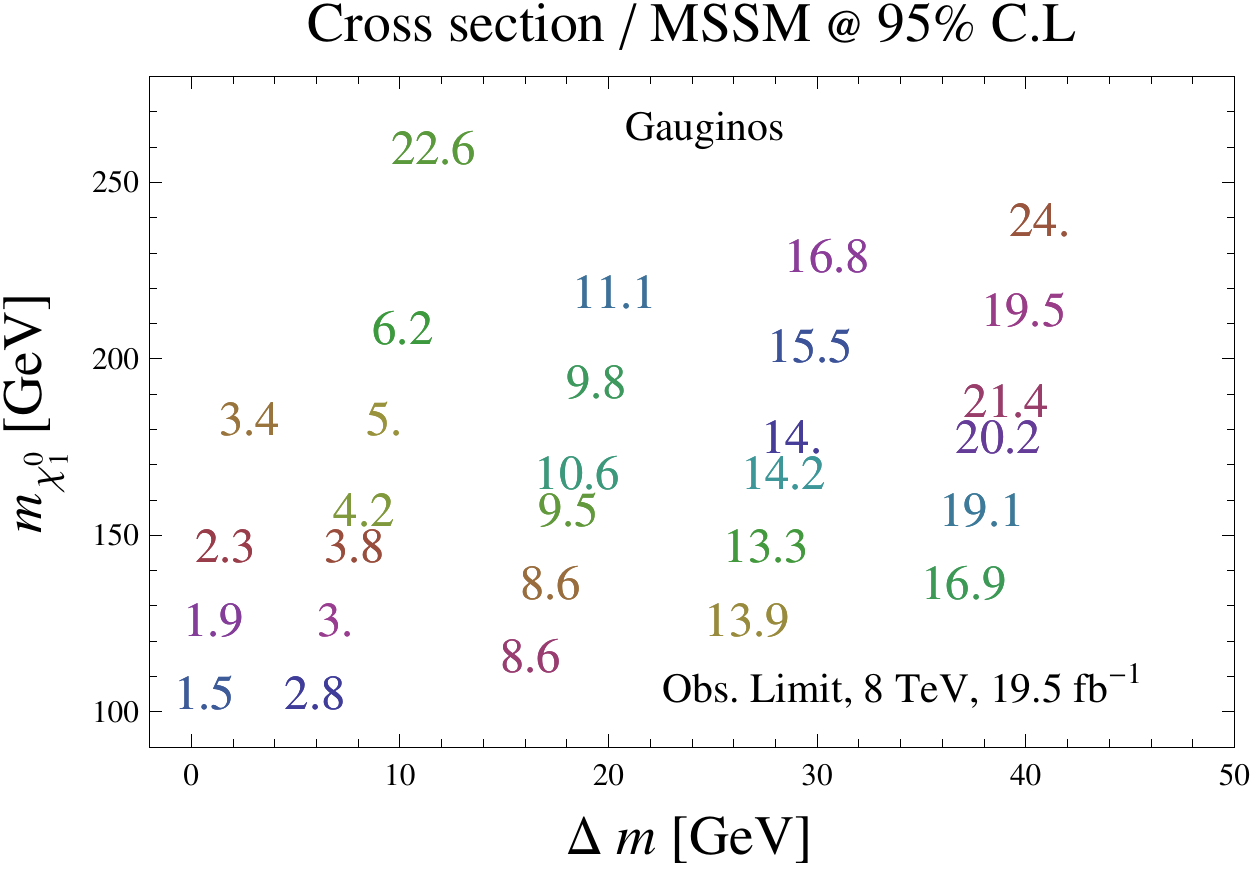}
\newline
(a)
\end{center}
\end{minipage}
\hspace{0.5cm}
\begin{minipage}[b]{0.47\linewidth}
\begin{center}
\includegraphics[width=1\textwidth]{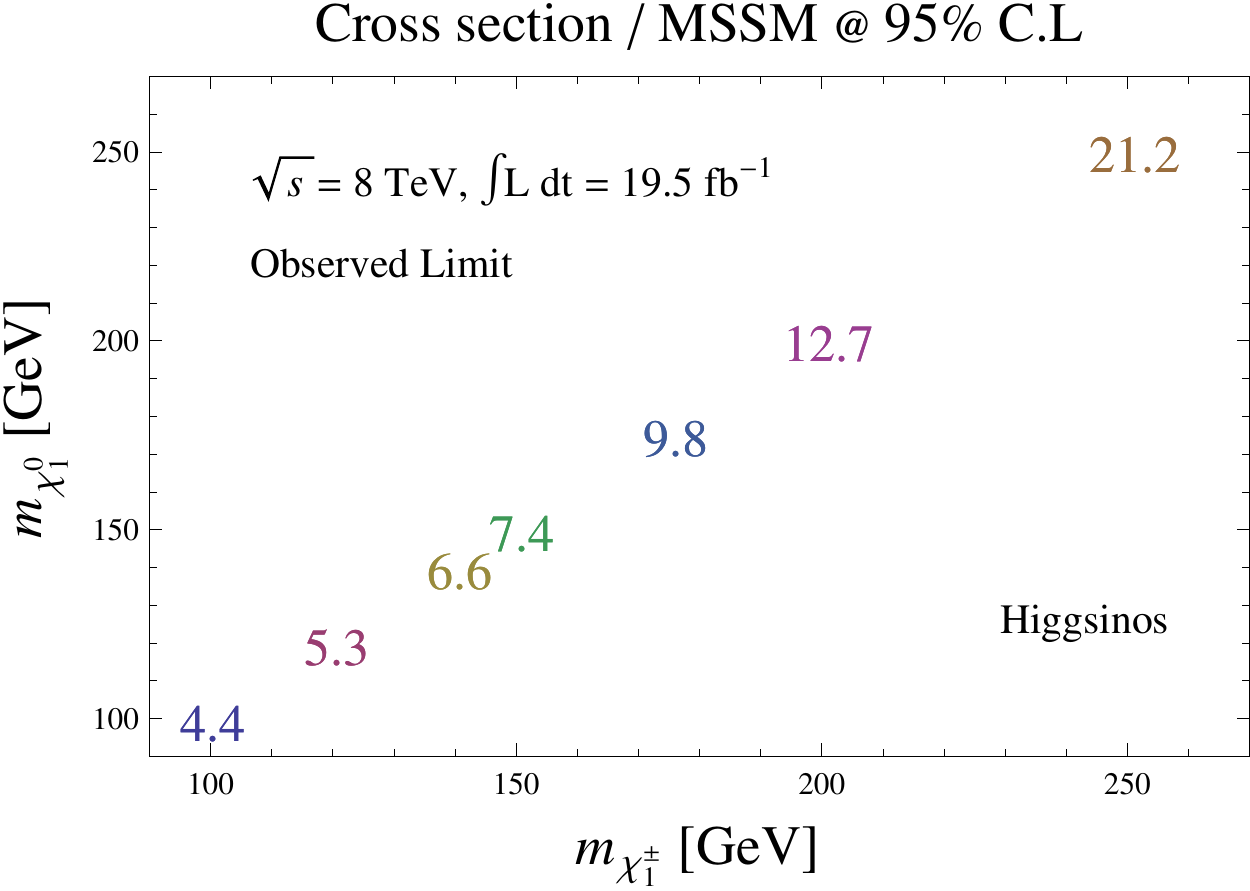}
\newline
(b)
\end{center}
\end{minipage}
\caption{Cross sections normalized to the MSSM value probed at the 95 \% C.L. ($\sigma = 2$) in the CMS analysis at the 8~TeV LHC (19.5 fb$^{-1}$) for the case of gauginos (left) and Higgsinos (right). For each point the number shown is equivalent to $2/\sigma$.}
 \label{fig:XSprobedgau}
\end{center}
\end{figure}

In Fig.~\ref{fig:XSprobedgau}a we present the cross section that can be probed at each point, normalized to the MSSM value for our signal point, for the gaugino case. To improve visibility we present the significances in the $\Delta m = \mchar - \mneut$ versus $\mneut$ plane. We see that the straightforward recast of the CMS analysis can only exclude cross sections larger than 1.5 times the MSSM result, and for heavier masses this number goes up to ${\cal O} (30)$. 
The corresponding results for the Higgsino case are shown in Fig.~\ref{fig:XSprobedgau}b. Since the Higgsino cross section is smaller, the sensitivity is even lower, and only models with a 4-20 times enhanced production rate can be probed. Reducing the systematic error does not seem to have a dramatic impact on these results, which at this point still seem to be limited by statistics.

We note, however, that the analysis can be improved in several ways. The inclusion of the NLO effects might enhance the signal enough to be sensitive to the MSSM, at least in the leftmost corner of the parameter space. One could also consider relaxing some of the previous cuts, in order to allow more signal events in the sample, and design an optimised analysis for compressed electroweakinos. 
Another way to improve the analysis is to exploit the presence of soft leptons (electrons or muons) in the sample. Here we define a \emph{soft muon} by the condition $ 5 ~{\rm GeV} < \ptmu < 20~{\rm GeV}$, and a \emph{soft electron} by $ 10 ~{\rm GeV} < \ptel < 20~{\rm GeV}$. The lower value of the transverse momentum is chosen according to the CMS capabilities of identifying muons~\cite{Chatrchyan:2012xi} and electrons~\cite{CMS:2010bta} with a high efficiency\footnote{These values could be lowered to 3 and 5 GeV respectively, but with an identification efficiency of about 20 \%. A better understanding of soft leptons at the LHC could lead to an important improvement in the sensitivity for this search. }.

  We note that due to the lepton veto applied in the CMS analysis, the only soft leptons present in the sample are muons with $5 ~{\rm GeV}  < \ptmu < 10~{\rm GeV}$. Hence, right after the CMS lepton veto, events with at least one soft muon account from 0.5\%~-~8\% of the total number of events, and given the limited statistics of the 8 TeV dataset the analysis will not improve significantly. Using our definition of soft leptons not only one gains more events, but also more events with soft-leptons, since now the soft-lepton fraction ranges from 1\% to 37\%. Thus, relaxing the lepton veto could be a good option to enhance the current sensitivity. However, statistics is still a limiting factor for the 8 TeV dataset, and we will consider a relaxed lepton veto for the 14 TeV LHC case.
  
  

We conclude that the 8 TeV data can not probe the MSSM cross sections at this point. If the NLO corrections for the signal processes would turn out to be large, if the lepton efficiencies for low $p_T$ were improved, it might  be possible to get sensitivity to very light and degenerate gauginos~\footnote{If $\beta = 1 \%$ and a 20 \% improvement on the sensitivity were to be achieved, then $\mneut \sim 125~\rm{GeV}, \Delta m \sim 1~\rm{GeV}$, and $\mneut \sim 100~\rm{GeV}, \Delta m \sim 5~\rm{GeV}$ could be tested at the $2-\sigma$ level.}. Higgsinos instead seem to be out of reach at 8~TeV. 
\section{14 TeV projection}
\label{sec:14TeV}
Here we present an estimate of the sensitivity of an optimised monojet search for almost degenerate electroweakinos at the 14~TeV LHC with 300~fb$^{-1}$ and at the high luminosity LHC (HL-LHC) with 3000~fb$^{-1}$. 

Signals and backgrounds are generated as described in the previous sections. For the 8~TeV analysis the background normalisation for the dominant $(Z\to\nu\nu)+{\rm jets}$ and $(W \to \ell\nu_\ell)+{\rm jets}$ backgrounds was obtained from data using control regions. Comparing with our simulation of the backgrounds, we find that correction factors of 0.83 and 1.01 would have to be applied to our Monte Carlo results to accurately model the backgrounds. Since these numbers are very close to unity, we will not apply any K-factors to these two backgrounds for our 14~TeV analysis. This should correspond to a less than 20\% error on our sensitivity estimates. The $t\bar{t}$ cross section increases notably at 14~TeV, and is subject to large radiative corrections. Therefore we normalise our $t\bar{t}$ event sample to the inclusive cross section of 939 pb obtained using HATHOR~\cite{Aliev:2010zk}.

The general direction of the analysis will be similar to the 8~TeV case, however we will perform several modifications to improve the sensitivity beyond that of a pure monojet search. 
First, to account for the higher center-of-mass energy, we will increase the preselection cuts to $\met > 300$~GeV and $\ptjone > 300$~GeV. Furthermore we will increase the threshold for the lepton vetoes to $\ptl > 20$~GeV for $\ell=e,\mu$, while keeping the tau veto at the same value, according to our definition of a soft lepton. 
We will perform two separate analyses:

\textbf{Pure monojet:} In analogy with the 8~TeV analysis, we will define signal regions with successively stronger $\met$ cuts, $\met > 300, 350,\dots,1000$~GeV, and derive the sensitivity $\sigma$ from the most significant bin for each signal point. Note that different from the 8~TeV case, we impose symmetric cuts on $\met$ and $\ptjone$, i.e. each bin is defined by an $\ptcut$ such that $\met > \ptcut$ and $\ptjone > \ptcut$. Such a symmetric cut is already used in the ATLAS 8~TeV analysis~\cite{ATLAS:2012ky}. The sensitivity can be further enhanced (by about 10-15 \%) by performing one or two additional cuts. The first cut is to veto on the second jet of the event, that is, discarding events that fulfill $\ptjtwo > 100$ GeV with $|\eta(j_2) |\leq 2$~\cite{Han:2013usa}. 
The second cut is a veto on events with reconstructed soft taus, and is only applied for highly degenerate spectra ($\Delta m \le 3.5 $ GeV), including all Higgsino signal points. For larger mass gaps the $\tau$ veto leads to an slight decrease of the significance, and hence we do not apply it. For clarity reasons, all the cuts used are shown in Table~\ref{tab:cuts}.

\begin{table}[tdp]
\begin{center}
\begin{tabular}{|c|c|c|}
\hline
Cuts & 8 TeV & 14 TeV \\
\hline
Preselection & & \\
\hline
$\ptjone $ (within $|\eta|<2.4$)  larger than & ~~120~GeV~~ & ~~300~GeV~~ \\
$\met$ larger than & 200~GeV & 300~GeV \\
$N_{jet} (p_T > 30~{\rm GeV}) \le 2$ & applied & applied \\
$\Delta \phi (j_1, j_2) < 2.5$ & applied & applied \\
veto on $e,\mu$ with $p_T$ larger than  & 10~GeV & 20~GeV \\
veto on $\tau$ with $p_T$ larger than  & 20~GeV & 20~GeV \\
\hline
Analysis & & \\
\hline 
$\met$ (50~GeV steps) & 250~GeV - 500~GeV &  \\
$\met \geq \ptcut$, $\ptjone \geq \ptcut $ (50~GeV steps) &  & 300~GeV - 1000~GeV \\
$\ptjtwo < 100$ GeV within $|\eta(j_2) |\leq 2$ & not applied & optional \\
$N(\tau) = 0$ & not applied & optional \\
\hline
\end{tabular}
\end{center}
\caption{Comparison of cuts used in 8~TeV and 14~TeV analyses. Cuts labelled as preselection are applied to all events. From the analysis cuts for each signal point the combination that maximizes the significance is chosen. }
\label{tab:cuts}
\end{table}%

\textbf{Soft leptons:} For this analysis, we divide the events, after preselection cuts, into exclusive bins with exactly zero, one or two soft leptons. Then for each bin with $i$ leptons we find the best sensitivity $\sigma_i$ for each signal point as in the monojet analysis, i.e. using successively stronger $\met$ and $\ptjone$ cuts, and add them in quadrature to obtain $\sigma^2 = \sigma_0^2 + \sigma_1^2 + \sigma_2^2$. This binning makes use of the fact that almost degenerate electroweakinos can still produce soft leptons in their decays, and we therefore expect that this will improve the sensitivity for MSSM inspired scenarios, whereas the pure monojet analysis is optimised for isolated dark matter candidates. In the 0-lepton and 1-lepton bin the cuts are similar to the ones used in the monojet analysis The 2-lepton bin has fewer events (${\cal O}(10-100)$ signal and ${\cal O}(2000)$  background for ${\cal L} =  300~\rm{fb}^{-1}$), and the best cut depends on the specific signal point. Here one can also gain sensitivity for the highly degenerated spectrum by splitting the events according to the lepton flavor. The background has $40 \%, 35 \%$ and $25 \%$ for the $\mu e$, $\mu \mu$ and $ee$ bins, but for the signal these numbers vary with $\Delta m$. For instance, in the Higgsino case, as well as for highly squeezed spectra, one typically has (20,70,10)\% for the same occupancy fractions, the reason being that for highly squeezed spectra the leptons become increasingly softer, and the muons are reconstructed down to lower momenta.

As we have seen in the previous section, systematic errors must not be neglected in these analyses. For our projections, we will consider a pessimistic scenario with $\beta = 5\%$ and an optimistic scenario with $\beta = 1\%$. Here it is worth noting that in some signal regions in the ATLAS and CMS monojet searches the systematic error is already at the 3\%-4\% level.~\footnote{While a systematic error of 1\% appears attainable in the mono-jet analysis, it might turn out to be larger in the bins with soft leptons, due to  extra sources of systematic errors (fake rates, hadron decays, etc). This could slightly reduce the projected significance for some signal points. } 

To highlight the differences between the two analyses, and to emphasise the benefit of performing the soft lepton analysis, we will now discuss the limits expected for a few of the signal points in detail. In Tab.~\ref{tab:details} we show the sensitivity of each analysis  for (A) a model with highly degenerate spectrum, (B) a model with moderate splitting close to the LEP limit, (C) a model with relatively large mass gap, and (D), (E) two Higgsino points. Since different bins can give the best sensitivity for different signal points, also the number of background events will vary for each point. For the table we take a systematic error of $\beta = 1\%$, but we also report the number of signal and background events for 300 fb$^{-1}$, such that it is easy to estimate the significance with a different systematic error and/or luminosity.

It is instructive to first consider (A), where the electroweakinos are maximally degenerate. As expected, in this case very few soft leptons are reconstructed, and they only marginally contribute to the overall significance. The difference between a monojet and a soft-lepton analysis is thus small. This model has the largest cross section and a good acceptance, and is one of the few where a $5\sigma$ discovery would be possible with 300~fb$^{-1}$, provided $\beta=1\%$ can be reached. 

When the mass gap gets larger, the fraction of reconstructed soft leptons increases. In (B) we see that the monojet analysis would not allow discovery ($\sigma = 4.0$), but the addition of the soft lepton bins is crucial to increase the significance by 40 \%, thus allowing for discovery. We also note that the value of $\ptcut$ in the 1- and 2-lepton bins is slightly lower than in the monojet and the 0-lepton case, due to the fact that those bins have fewer events, and hence are more dominated by statistics. When the mass splitting becomes even larger (C), our monojet inspired analysis becomes inefficient. The reason is that in this regime, the jets and leptons from the off-shell $W$ and $Z$ decays become more energetic, and many events fail to pass the multijet or lepton vetoes. 
For such a  $\Delta m$, a better option is the boosted trilepton analysis of  Ref.~\cite{Gori:2013ala}. Indeed, (C) could even be excluded with the current 8-TeV dataset.

\begin{table}[htdp]

\begin{center}
\begin{tabular}{lc|c|c|c|c|c|c|c|}
\hline
 Point \# &  & A & B & C & D & E  \\
\hline
$M_1$  & (GeV)   & 107   & 100    & 100 & 1000 & 1000  \\
$M_2$  & (GeV)   & 100   & 100    & 130& 1000 & 1000  \\
$\mu$   & (GeV)   & 1000 & 1000  & 1000  & 100 & 250  \\
$m_{\chi^{\pm}}$ & (GeV) & 104.5 & 104.5 & 135.8  & 100.5 & 251.3  \\
$m_{\chi^0}$   & (GeV)      & 103.8 & 98.5 & 98.7       & 97.0   & 247.8  \\
\hline
Analysis &  &  &  &  &  &   \\
\hline
		 		 & S & 2654     &2327     &  331     & 1154     & 114  \\
 Monojet			 & B  & 31216 & 53230  & 53220  & 50656  & 12634  \\
	               & $\sigma$  & 7.4       & 4.0        & 0.6        &  2.1       & 0.7  \\
	                 & $\ptcut$  & 550      & 500      & 500       & 500       & 700  \\		
	   &$j_2 / \tau$-veto  & y/y       & y/n      & y/n       & y/n        & n/y  \\		

%
		\hline
		 		& S & 2555    & 1972   & 284      & 1071    & 109  \\
0-lepton     		& B  & 28648 & 47481 & 47481 & 47481 & 12058  \\
                        & $\sigma$ & 7.7       & 3.8       & 0.5       & 2.1       & 0.7  \\
 	                 & $\ptcut$ & 550      &  500       &  500       & 500        & 650  \\
            &$j_2 / \tau$-veto  & y/y        & y/n      & y/n       & y/n        & y/n  \\	

\hline
			 	& S  & 75      & 502   & 101       & 210      & 11  \\
 1- lepton 			& B  & 1433 & 9836 & 13885  & 10341  & 1433  \\
 	              & $\sigma$  & 1.9     & 3.6     & 0.6       & 1.4        & 0.3  \\
  	               & $\ptcut$   & 650      &  450     & 450  & 450      & 650  \\
	  &$j_2 / \tau$-veto  & n/y        & y/n      & n/n       & y/y      & n/y  \\	

\hline
				 & S & 18     & 39           & 39      & 58      & 6  \\
2-lepton  			 & B & 487   & 340        & 2320  & 560    & 560  \\
                        & $\sigma$ & 0.8     & 2.1          & 0.7     & 2.4     & 0.3  \\
 	                 & $\ptcut$ & 400    & 300 & 300    &  400  &  300   \\
            &$j_2 / \tau$-veto  & n/y     & n/n          & n/n     & n/y     & y/n  \\	
            &$\ell$  & all     & 2 $\mu$          & all     & all     & $2 \mu$  \\	
\hline
0+1+2 leptons  & $\sigma$ & 7.9  & 5.6 & 1.1 & 3.4 & 0.8  \\
\hline
\end{tabular}
\end{center}
\caption{Number of signal and background events for each analysis, for a total integrated luminosity of 300 fb$^{-1}$. The significance for each analysis is computed assuming a systematic error of $\beta = 0.01$. We also show the cuts that give the largest significance. }
\label{tab:details}
\end{table}%

Now we consider the Higgsino case, where the cross sections are smaller. In (D) we see not only that the soft lepton analysis allows for a 70\% increase of the significance (from 2.1 to 3.4), but also that the 2-lepton bin is more sensitive than the monojet analysis. Since the number of events for 300 fb$^{-1}$ is still relatively small, the significance can not be enhanced by performing additional cuts on this bin. This point is useful to illustrate the crucial role of $\beta$ in the estimation of the sensitivity. For $\beta = 5 \%$ one would find $\sigma = 0.5$, and for $\beta = 0$ one would have $\sigma = 5.1$. This also shows that discarding systematics errors in this kind of analysis is an unjustified assumption.

When the Higgsino mass gets larger (E) the main features of (D) remain the same, but with the signal being reduced by an order of magnitude, and hence such a point would be very hard (if not impossible) to test at the LHC. 

\begin{figure}[!htp]
\begin{center}
\begin{minipage}[b]{0.47\linewidth}
\begin{center}
\includegraphics[width=1\textwidth]{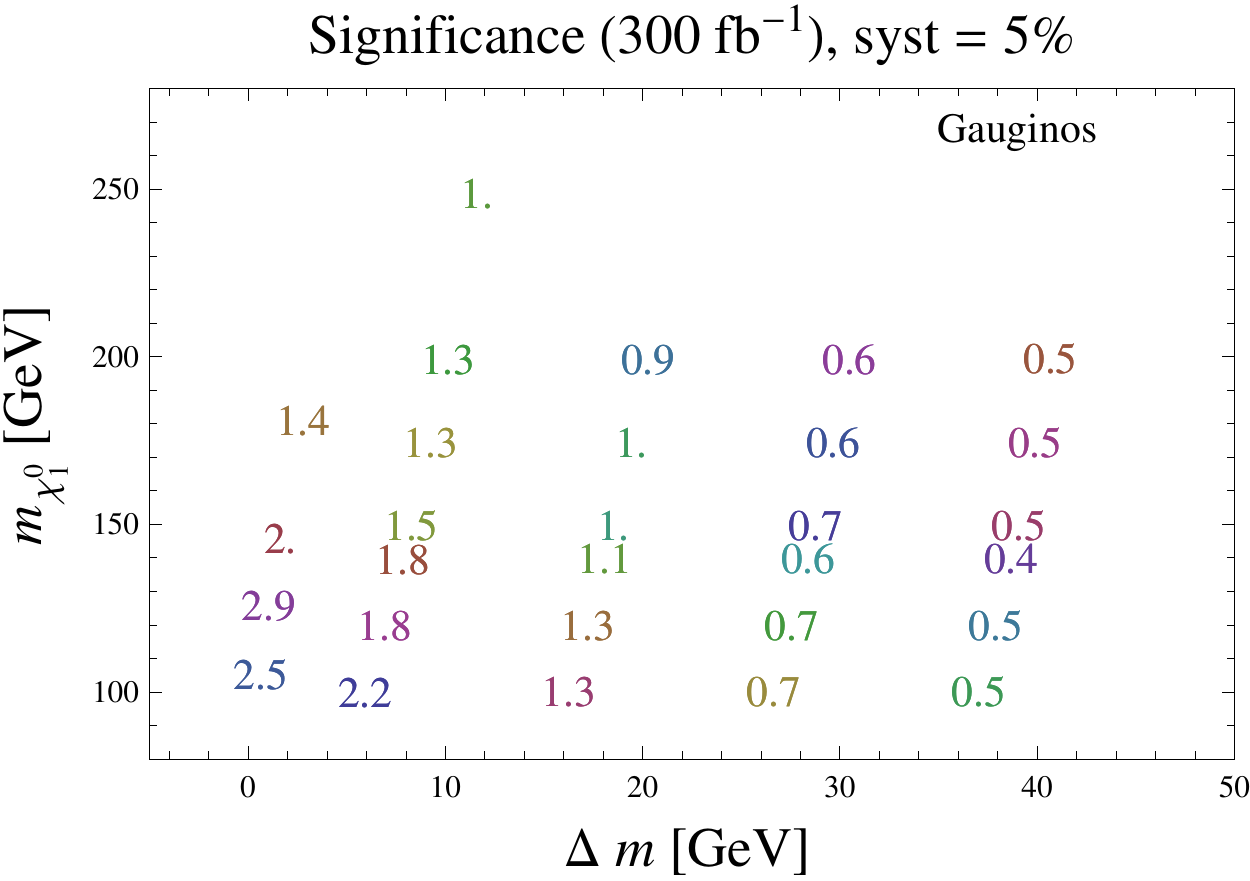}
\newline
(a)
\end{center}
\end{minipage}
\hspace{0.5cm}
\begin{minipage}[b]{0.47\linewidth}
\begin{center}
\includegraphics[width=1\textwidth]{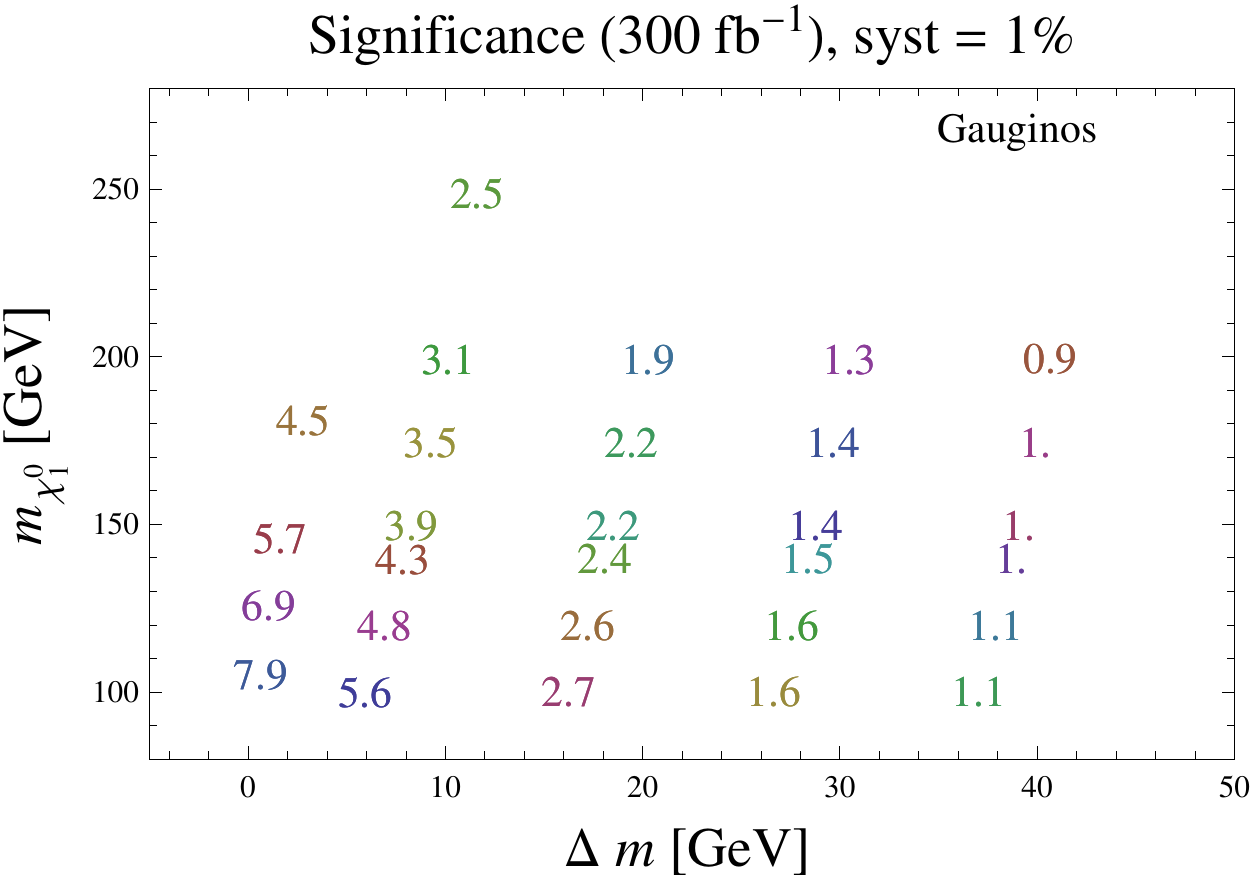}
\newline
(b)
\end{center}
\end{minipage}
\caption{Significance at the 14 TeV LHC, with 300 fb$^{-1}$, considering a systematic error of (a) 5\% and (b) 1\%, for the gaugino case. Here $\Delta m = m_{\chi_1^\pm} - m_{\chi_1^0}$. }
 \label{fig:14gau300}
\end{center}
\end{figure}

\begin{figure}[!htp]
\begin{center}
\begin{minipage}[b]{0.47\linewidth}
\begin{center}
\includegraphics[width=1\textwidth]{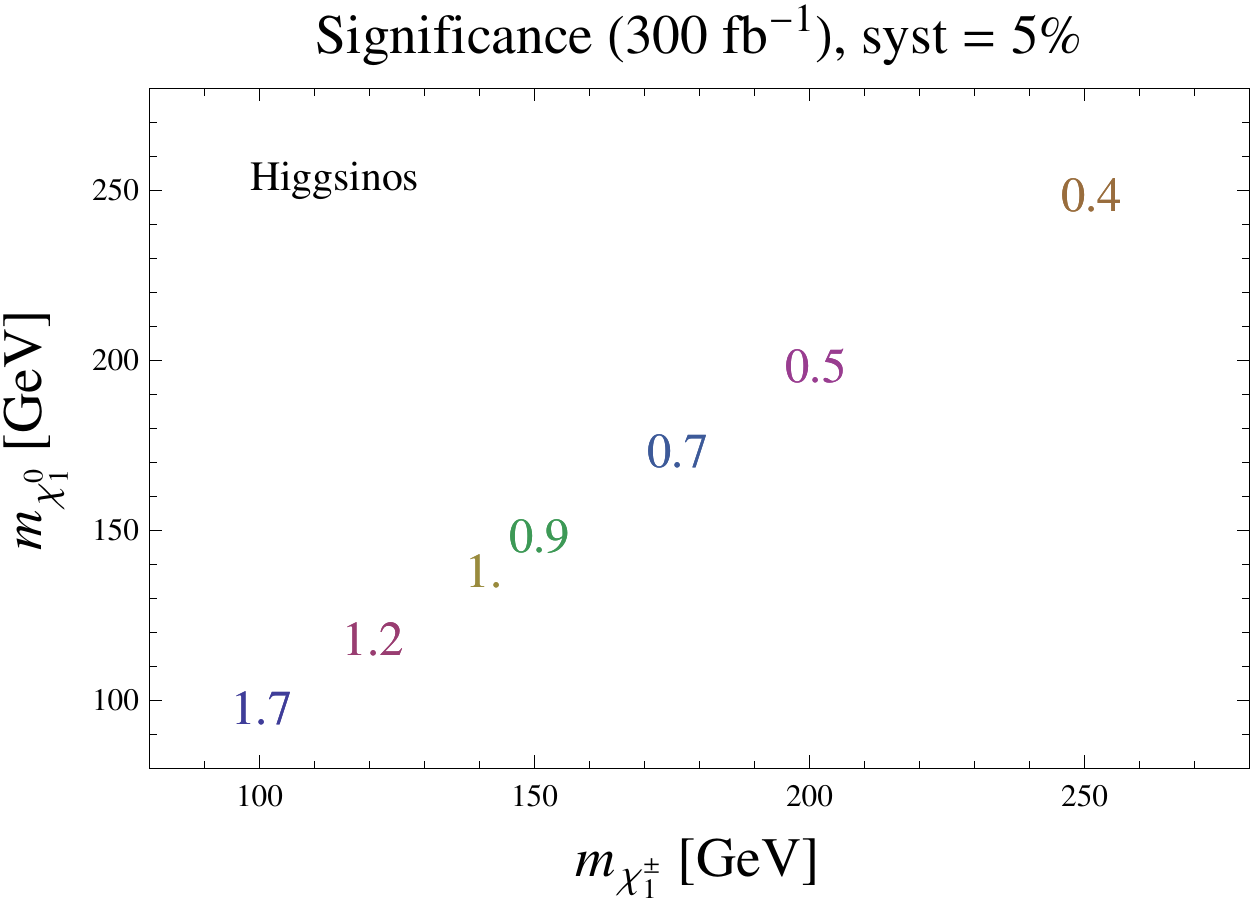}
\newline
(a)
\end{center}
\end{minipage}
\hspace{0.5cm}
\begin{minipage}[b]{0.47\linewidth}
\begin{center}
\includegraphics[width=1\textwidth]{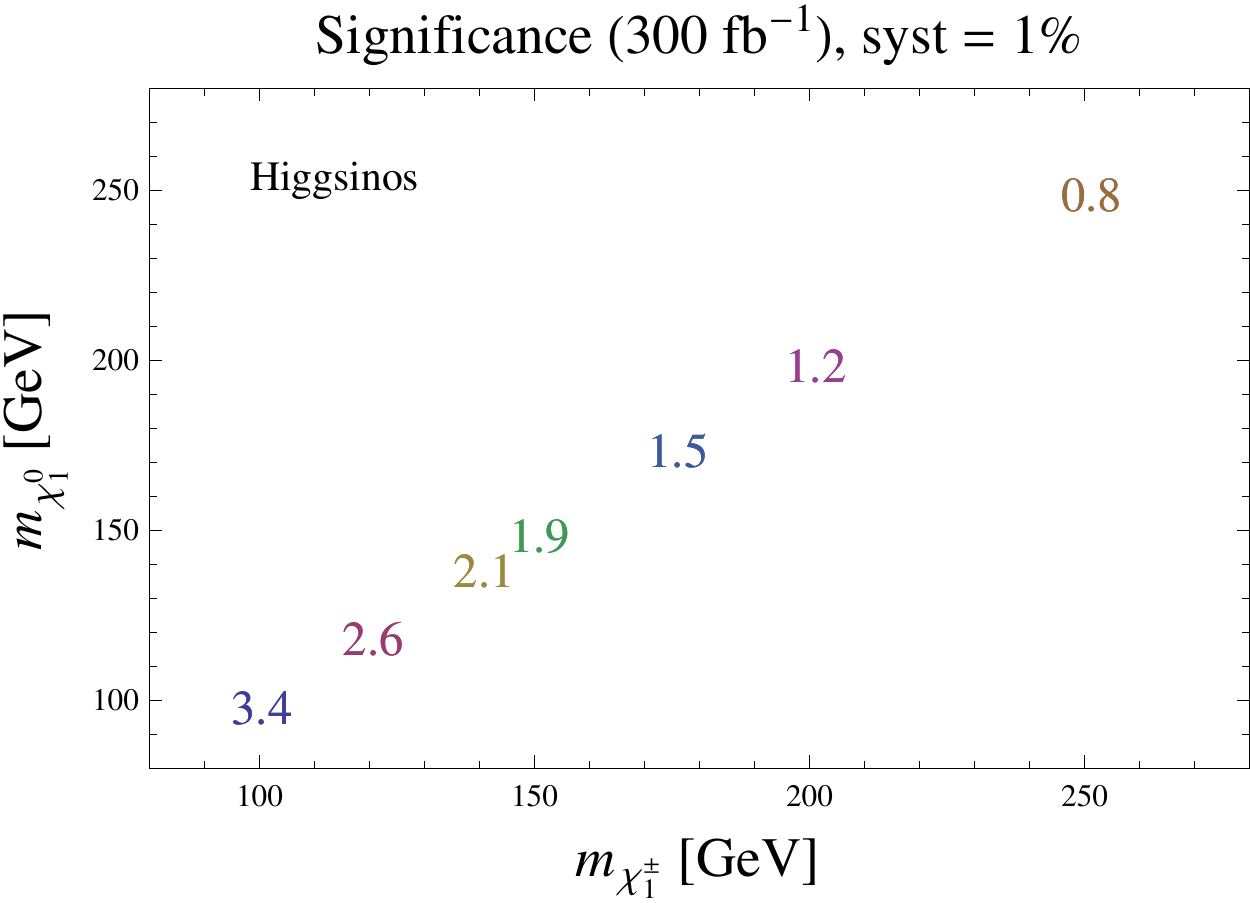}
\newline
(b)
\end{center}
\end{minipage}
\caption{Significance at the 14 TeV LHC, with 300 fb$^{-1}$, considering a systematic error of (a) 5\% and (b) 1\%, for the Higgsino case.}
 \label{fig:14hig300}
\end{center}
\end{figure}

\begin{figure}[!htp]
\begin{center}
\begin{minipage}[b]{0.47\linewidth}
\begin{center}
\includegraphics[width=1\textwidth]{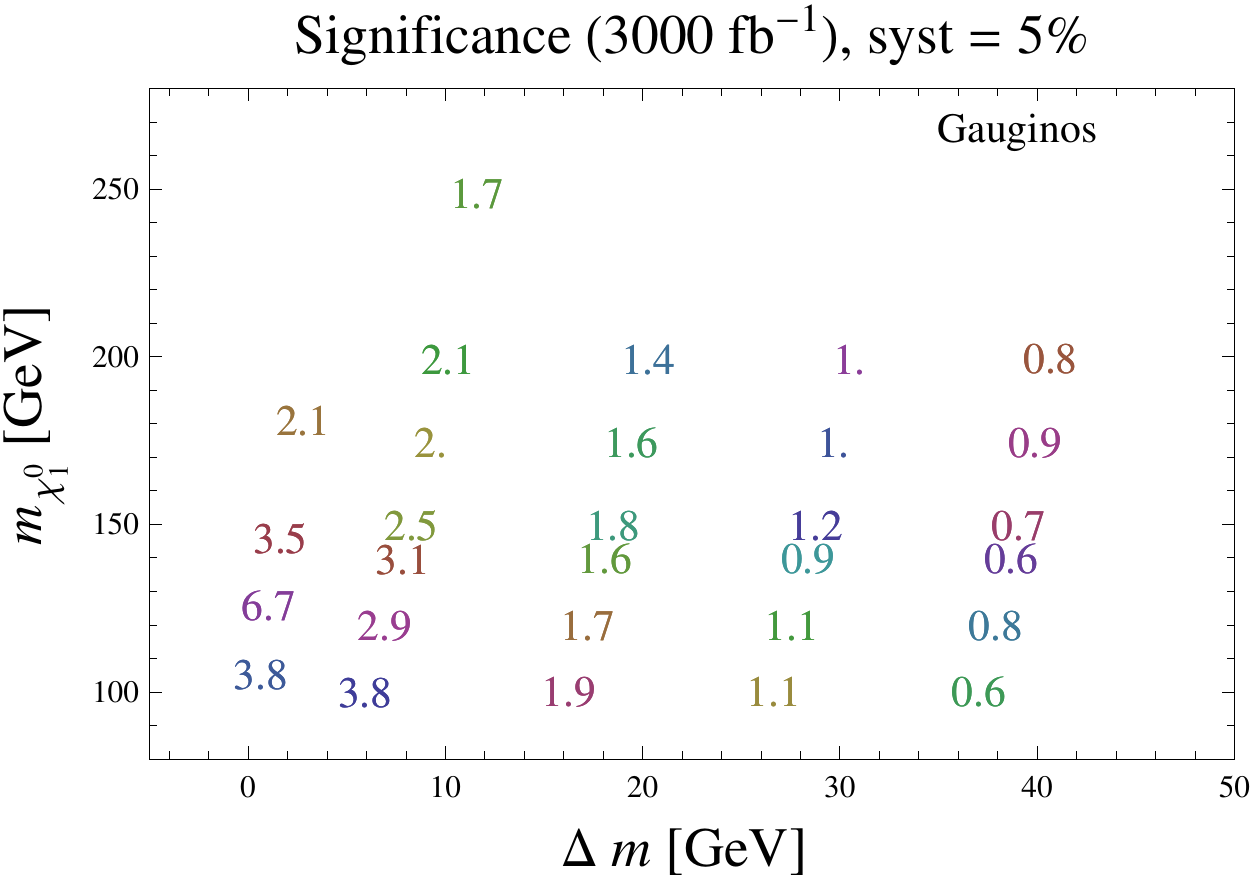}
\newline
(a)
\end{center}
\end{minipage}
\hspace{0.5cm}
\begin{minipage}[b]{0.47\linewidth}
\begin{center}
\includegraphics[width=1\textwidth]{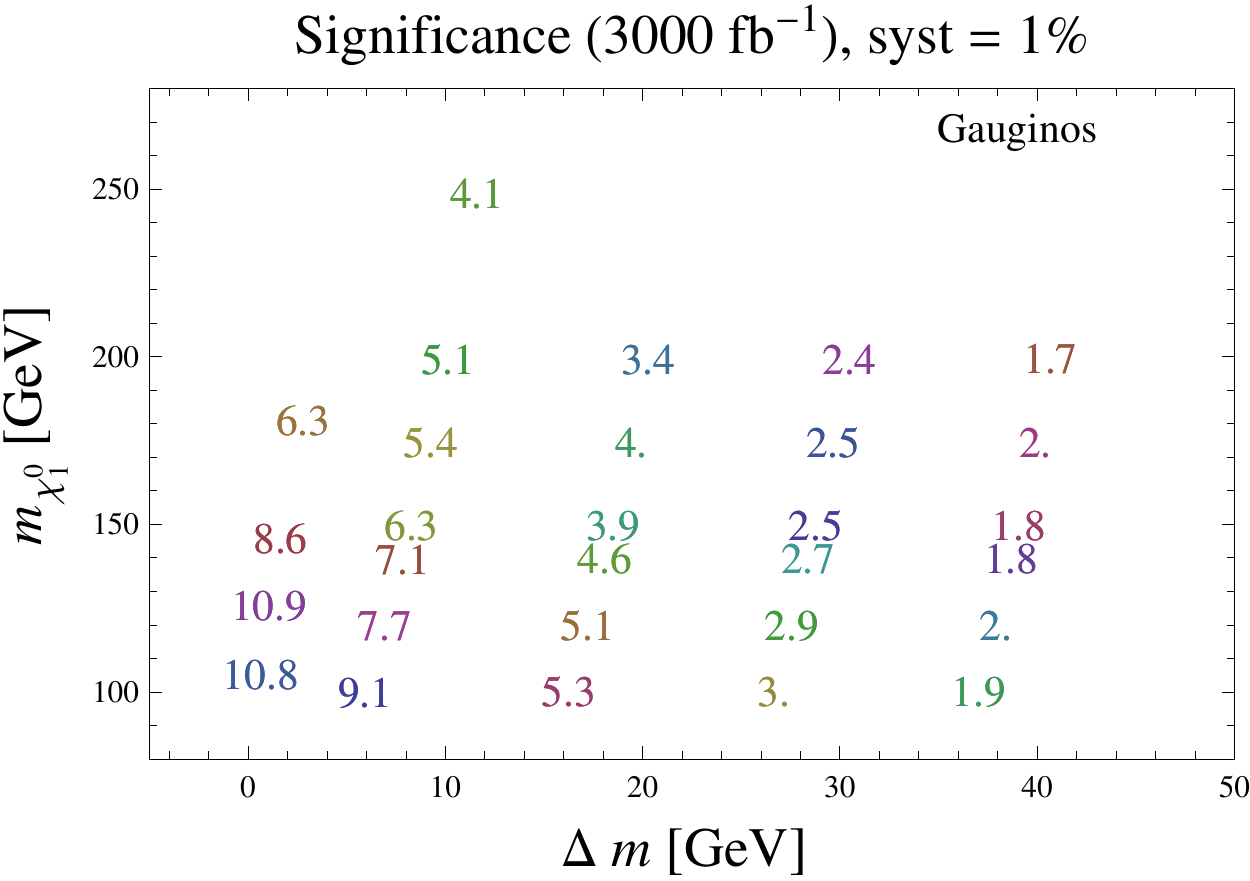}
\newline
(b)
\end{center}
\end{minipage}
\caption{Significance at the 14 TeV LHC, with 3000 fb$^{-1}$, considering a systematic error of (a) 5\% and (b) 1\%, for the gaugino case.}
 \label{fig:14gau3000}
\end{center}
\end{figure}

\begin{figure}[!htp]
\begin{center}
\begin{minipage}[b]{0.47\linewidth}
\begin{center}
\includegraphics[width=1\textwidth]{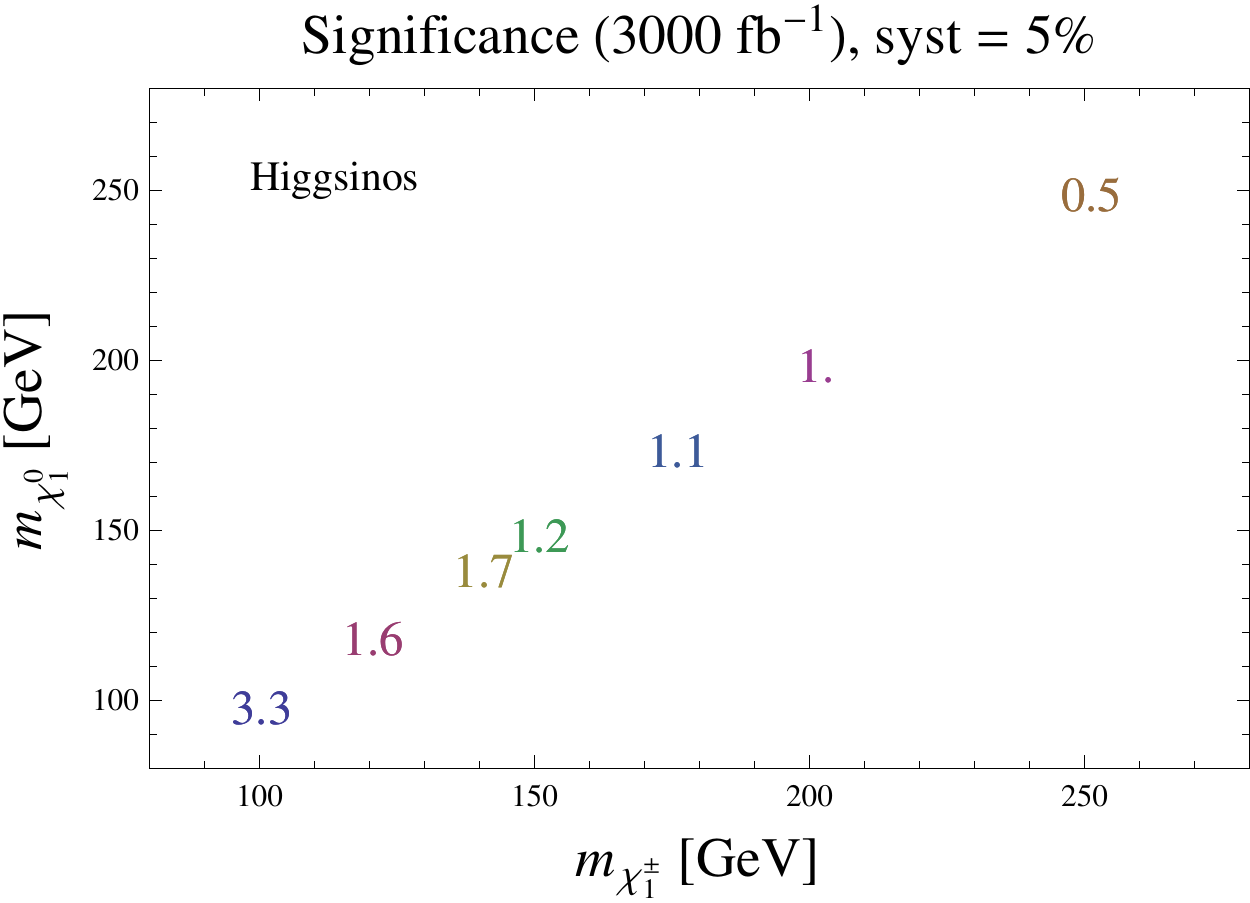}
\newline
(a)
\end{center}
\end{minipage}
\hspace{0.5cm}
\begin{minipage}[b]{0.47\linewidth}
\begin{center}
\includegraphics[width=1\textwidth]{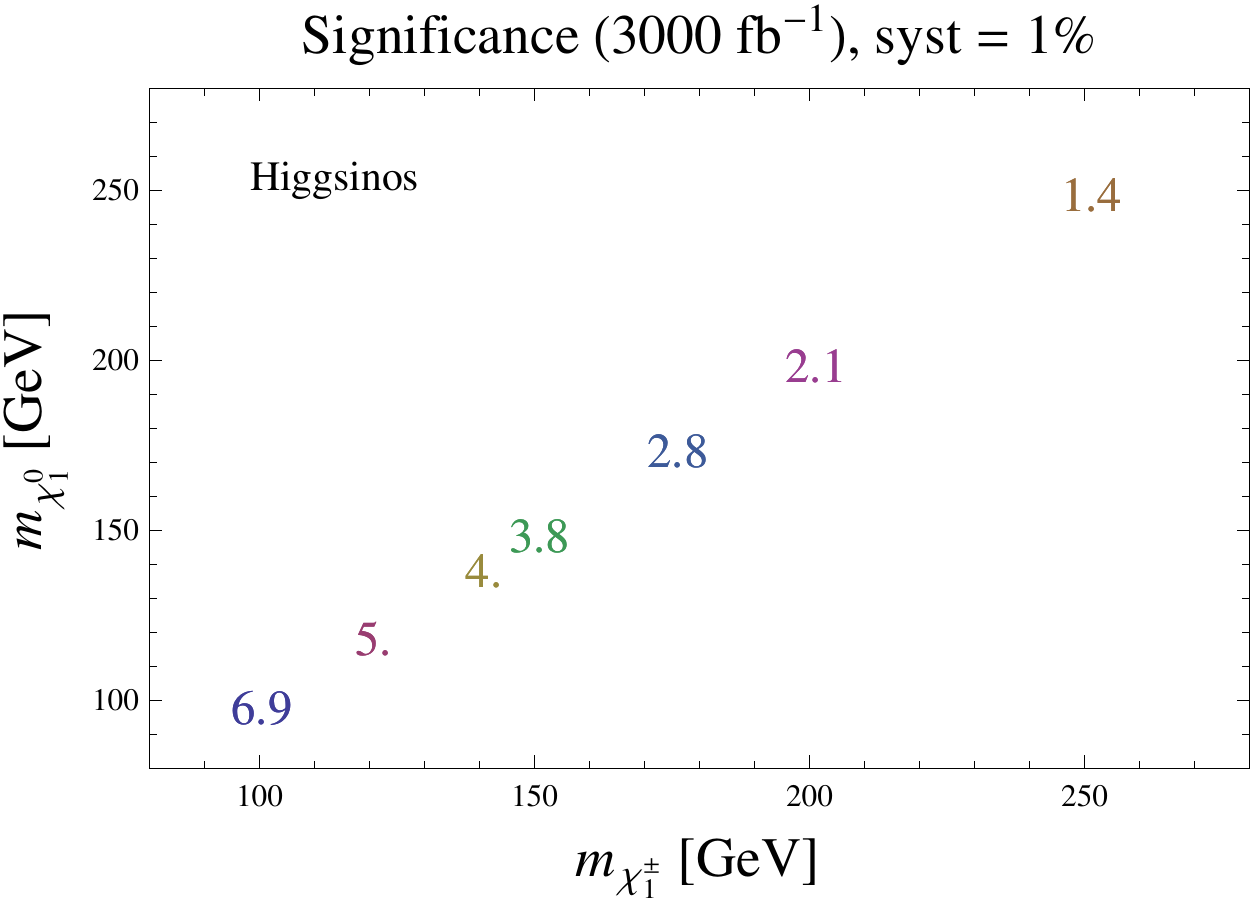}
\newline
(b)
\end{center}
\end{minipage}
\caption{Significance at the 14 TeV LHC, with 3000 fb$^{-1}$, considering a systematic error of (a) 5\% and (b) 1\%, for the Higgsino case.}
 \label{fig:14hig3000}
\end{center}
\end{figure}

Our final results for the gaugino case are shown in Fig.~\ref{fig:14gau300} and Fig.~\ref{fig:14gau3000}, for total integrated luminosities of  300 fb$^{-1}$  and 3000 fb$^{-1}$ respectively. Fig.~\ref{fig:14hig300} and Fig.~\ref{fig:14hig3000} are the analogous for the Higgsino case. For each case we show two results corresponding to systematic errors of 5\% (left panels) or 1\% (right panels). The shown significances correspond to the most sensitive analysis for each point, which always is the combined soft lepton analysis. 

From the previous Figures we see how the level of systematics strongly limit the exclusion reach. Indeed, with $\beta = 5 \%$ and 300 fb$^{-1}$ one can only probe the highly compressed spectrum with $\Delta m \sim 1-5$~GeV, and even multiplying the luminosity by 10 will only extend the reach up to 10 GeV for $\Delta m$, but also to higher masses. We note that for these compressed spectra we are in the situation of point (A), where the 0-lepton bin dominates, and the soft leptons do not dramatically increase the sensitivity.

If one decreases $\beta$ down to 1 \% the situation is improved dramatically. For gauginos, one can discover the highly compressed spectra, and obtain evidence (exclusion) for \emph{medium} mass gaps of 10 (20) GeV. For Higgsinos one can test masses up to 150 GeV. Having 3000 fb$^{-1}$ with $\beta = 1 \%$ would allow to discover many points, and cover almost all of the parameter space. We note that for these intermediate values of $\Delta m$ the addition of the soft leptons is crucial to be able to have these points within LHC reach.

For the points with $\Delta m \sim 35 ~\rm{GeV}$, which seem to be difficult to probe even in the most optimistic scenario, one should note that they are in principle in the reach of the analysis of Ref.~\cite{Gori:2013ala}, which claims to probe mass gaps down to $\Delta m = 12$ GeV at the 14~TeV LHC. Hence we find a nice complementarity between the monojet+soft lepton search and the work of Ref.~\cite{Gori:2013ala}, which becomes inefficient for smaller mass gaps. Eventually the few points in between could be tested (or even discovered) by combining both strategies.
%
\section{Conclusions}
\label{sec:conclu}

Monojet searches have been used by the LHC experiments to probe and constrain simplified dark matter models, where the DM candidate is coupled to the SM via effective operators or through very heavy mediators. Here we apply a monojet inspired search to a more complex scenario with light charginos and neutralinos. Using the MSSM as an example, we evaluate the sensitivity of the existing monojet searches to light degenerate gaugino and Higgsino scenarios, and propose improved search strategies to probe these scenarios at the 14~TeV LHC. 

Conventional searches for direct chargino and neutralino production at the LHC fail in the limit where the mass gaps between the produced particles are small and the multilepton and $\met$ triggers fail to pick up the event. An additional hard jet from ISR can recoil against the neutralinos and lead to a visible event with a hard jet and large $\met$. Furthermore soft leptons produced in chargino or neutralino decays can be boosted above the reconstruction thresholds and become visible again. 

We have performed a careful reanalysis of the 8~TeV CMS monojet search to validate our Monte Carlo setup, and we find that while the existing search is not sensitive to directly produced charginos and neutralinos, scenarios with a few times enhanced cross sections are in reach of the existing monojet search. We further discuss several improvements that could be done to optimise the existing analysis to MSSM-like scenarios. 

For the 14~TeV LHC, we perform several optimisations to improve the sensitivity of the search, and compare with the results of a pure monojet analysis. In particular, we find that the sensitivity of the monojet search can be sizeably enhanced by the addition of bins with soft leptons. The enhancement is small if the spectrum is very squeezed ($\Delta m < 5$ GeV), but if the mass splitting is between 10 and 30 GeV then the soft leptons are crucial to probe signal points in parameter space. In those cases the monojet search rapidly loses steam, and the bins with soft leptons can yield more sensitivity than just considering events without leptons. For larger gaps ($\Delta m \gtrsim 35~\rm{GeV}$) the search becomes less efficient, since reconstructed jets and leptons tend to have larger transverse momenta and thus more signal events fail the monojet cuts. For such scenarios other search strategies exist. 

We find that the 14 TeV LHC with 300 (3000) fb$^{-1}$ and with a 1\% systematic error can exclude gaugino masses up to 250 GeV for mass splittings below 10 (40) GeV, while for Higgsinos one can cover up to 150 (200) GeV.  For the gaugino case, $\Delta m < 15$ GeV can be discovered for chargino masses up to 250 GeV, while for Higgsinos discovery is only possible if the chargino is lighter than 125 GeV. 

The sensitivity drops significantly when a larger systematic error is assumed. It is therefore important to include these errors in the estimates. Given the current level of systematic errors and the possibility for improvements with more statistics, we are confident that the experiments can reach the 1\%-5\% range used for our analysis, and maybe even go beyond that. 

Several aspects of the analysis could be improved to further optimise the monojet search for MSSM-inspired scenarios. First, one should see if the third jet veto and lepton vetoes can be relaxed, in order to improve the sensitivity for moderate mass splittings of 20-40~GeV. Also the thresholds for lepton reconstruction should be further relaxed, if possible. 

Finally we want to stress that the search can also be applied to other BSM scenarios that involve weakly coupled multiplets, for example models with vectorlike leptons~\cite{Joglekar:2012vc,ArkaniHamed:2012kq} and models of mixed or coannihilating dark matter~\cite{Cheung:2013dua,Bell:2013wua}. For the vectorlike lepton case the sensitivity can easily be estimated when one recalls that a vectorlike lepton doublet has the same quantum numbers as a Higgsino. Models like those of~\cite{Joglekar:2012vc,ArkaniHamed:2012kq}, where in addition SU(2) singlets are present will then have a larger cross section, however when the multiplets are split e.g. by large Yukawa couplings the direct search for the lighter states using monojets will again be difficult. 

\acknowledgments 

We would like to thank the CERN Theory Division for the the lively atmosphere at the 2012 Summer Institute on LHC physics, where this work was originated. We would like to thank Yevgeny Kats for useful comments on the manuscript. J.Z would like to thank Ezequiel Alvarez, Rikkert Frederix, Benjamin Fuks, Barbara Jaeger and Andreas Papaefstathiou for useful discussions.

JZ is supported by the ERC Advanced Grant EFT4LHC of the European Research Council, the Cluster of Excellence Precision Physics, Fundamental Interactions and Structure of Matter (PRISMA-EXC 1098).
PS's work before September 2013 was supported by the U.S. Department of Energy, Division of High Energy Physics, under grant numbers 
DE-AC02-06CH11357 and DE-FG02-84ER40173.


\end{document}